\documentclass[aps,twocolumn,superscriptaddress]{revtex4-2}
\usepackage{amsfonts}
\usepackage{dcolumn}
\usepackage[left]{lineno}
\usepackage{bm}
\usepackage{tikz}
\usepackage[colorlinks=true,citecolor=blue,urlcolor=blue]{hyperref}
\usepackage{amsmath,amsfonts,amssymb,times,natbib}
\usepackage[standard]{ntheorem}

\begin{document}




\title{Realization of exceptional points along a synthetic orbital angular momentum dimension}
\author{Mu Yang \footnote{These authors contribute equally to this work\label{Contribute}}}
\thanks{These authors contribute equally to this work\label{Contribute}}
\affiliation{CAS Key Laboratory of Quantum Information, University of Science and Technology of China, Hefei 230026, People's Republic of China}
\affiliation{CAS Center For Excellence in Quantum Information and Quantum Physics, University of Science and Technology of China, Hefei 230026, People's Republic of China}

\author{Hao-Qing Zhang\textsuperscript{\ref{Contribute}}}
\affiliation{CAS Key Laboratory of Quantum Information, University of Science and Technology of China, Hefei 230026, People's Republic of China}
\affiliation{CAS Center For Excellence in Quantum Information and Quantum Physics, University of Science and Technology of China, Hefei 230026, People's Republic of China}

\author{Yu-Wei Liao\textsuperscript{\ref{Contribute}}}
\affiliation{CAS Key Laboratory of Quantum Information, University of Science and Technology of China, Hefei 230026, People's Republic of China}
\affiliation{CAS Center For Excellence in Quantum Information and Quantum Physics, University of Science and Technology of China, Hefei 230026, People's Republic of China}

\author{Zheng-Hao Liu}
\affiliation{CAS Key Laboratory of Quantum Information, University of Science and Technology of China, Hefei 230026, People's Republic of China}
\affiliation{CAS Center For Excellence in Quantum Information and Quantum Physics, University of Science and Technology of China, Hefei 230026, People's Republic of China}

\author{Zheng-Wei Zhou}
\affiliation{CAS Key Laboratory of Quantum Information, University of Science and Technology of China, Hefei 230026, People's Republic of China}
\affiliation{CAS Center For Excellence in Quantum Information and Quantum Physics, University of Science and Technology of China, Hefei 230026, People's Republic of China}
\affiliation{Hefei National Laboratory, University of Science and Technology of China, Hefei 230088, People's Republic of China}

\author{Xing-Xiang Zhou}
\affiliation{CAS Key Laboratory of Quantum Information, University of Science and Technology of China, Hefei 230026, People's Republic of China}
\affiliation{CAS Center For Excellence in Quantum Information and Quantum Physics, University of Science and Technology of China, Hefei 230026, People's Republic of China}
\affiliation{Hefei National Laboratory, University of Science and Technology of China, Hefei 230088, People's Republic of China}

\author{Jin-Shi Xu}\email{jsxu@ustc.edu.cn}
\affiliation{CAS Key Laboratory of Quantum Information, University of Science and Technology of China, Hefei 230026, People's Republic of China}
\affiliation{CAS Center For Excellence in Quantum Information and Quantum Physics, University of Science and Technology of China, Hefei 230026, People's Republic of China}
\affiliation{Hefei National Laboratory, University of Science and Technology of China, Hefei 230088, People's Republic of China}

\author{Yong-Jian Han}\email{smhan@ustc.edu.cn}
\affiliation{CAS Key Laboratory of Quantum Information, University of Science and Technology of China, Hefei 230026, People's Republic of China}
\affiliation{CAS Center For Excellence in Quantum Information and Quantum Physics, University of Science and Technology of China, Hefei 230026, People's Republic of China}
\affiliation{Hefei National Laboratory, University of Science and Technology of China, Hefei 230088, People's Republic of China}

\author{Chuan-Feng Li}\email{cfli@ustc.edu.cn}
\affiliation{CAS Key Laboratory of Quantum Information, University of Science and Technology of China, Hefei 230026, People's Republic of China}
\affiliation{CAS Center For Excellence in Quantum Information and Quantum Physics, University of Science and Technology of China, Hefei 230026, People's Republic of China}
\affiliation{Hefei National Laboratory, University of Science and Technology of China, Hefei 230088, People's Republic of China}

\author{Guang-Can Guo}
\affiliation{CAS Key Laboratory of Quantum Information, University of Science and Technology of China, Hefei 230026, People's Republic of China}
\affiliation{CAS Center For Excellence in Quantum Information and Quantum Physics, University of Science and Technology of China, Hefei 230026, People's Republic of China}
\affiliation{Hefei National Laboratory, University of Science and Technology of China, Hefei 230088, People's Republic of China}

\begin{abstract}
Exceptional points (EPs), at which more than one eigenvalue and eigenvector coalesce, are unique spectral features of Non-Hermiticity (NH) systems. They exist widely in open systems with complex energy spectra. 
We experimentally demonstrate the appearance of paired EPs in a periodical driven degenerate optical cavity along the synthetic orbital angular momentum (OAM) dimension with a tunable parameter. The complex-energy band structures and the key features of EPs, i.e. their Fermi arcs, parity-time symmetry breaking transition, energy swapping, and half-integer band windings are directly observed by detecting the cavity's transmission spectrum. Our results advance the fundamental understanding of NH physics and demonstrate the flexibility of using the photonic synthetic dimensions to implement NH systems.
 \end{abstract}

\date{\today}

\maketitle

\noindent{\bf \large Introduction}\\
The non-Hermiticity (NH) provides rich topology and unique physics distinct from the Hermitian system~\cite{NH}, among which the exceptional points (EPs)~\cite{EP,EP2} are prominent spectral features of the NH systems. 
EPs are branch singularities in the momentum space of an NH system, where two or more eigenenergies and eigenstates simultaneously coalesce and become degenerate. They have been observed in extensive nonconservative systems exchanging energies with the environment~\cite{spin2,acoustic0,acoustic1,EP2}. Many significant applications have been demonstrated for EP singularities, including the dramatical topological mode transport~\cite{switchE1,acoustic0}, fractional
topological charge measurement~\cite{halfcharge, crystal1} and ultrasensitive metrology~\cite{sensing2,sensing3}. 

In recent years, the NH systems with parity-time (PT) symmetry have caused special interests, in which the real energy spectrum is maintained as that promised by the Hermitian Hamiltonian~\cite{PT0,PT1}. Through spontaneous breaking, the eigenenergy spectrum becomes complex~\cite{PT2}. The PT symmetry breaking transition points are the companion EPs. 
The EPs have been observed in two- or higher- dimensional geometric parameter spaces including solid single spin systems~\cite{spin1,spin2}, acoustic cavities~\cite{acoustic1}, microwave cavities~\cite{microcavity1,microcavity2} or magnon polaritons~\cite{magnon}. One can conveniently tune the independent parameters in these systems to obtain the characteristic features of NH matter.

On the other hand, the EPs have been demonstrated in the real lattice such as photonic crystals~\cite{crystal2,crystal3} and waveguide arrays~\cite{waveguide1, waveguide2}. However, the physical dimensions are always not more than their geometric dimensions, and one of the approaches toward the high-dimensional NH system is to explore synthetic dimensions~\cite{Fan1}. High-dimensional physics, such as 4D quantum Hall physics~\cite{4D} and Weyl Physics~\cite{wely1,wely2}, can be studied even in low-dimensional real space. Moreover, some inspiring NH phenomena in synthetic lattices including windings and braid~\cite{Fan2, Fan3} of the complex-energy bands in synthetic frequency lattice, EPs in parity–time time-multiplexed lattice~\cite{time} have been investigated.

In this work, for the first time, we construct a NH system in a new configuration along the synthetic orbital angular momentum (OAM) dimension~\cite{OAM1,OAM2}. We directly measure the real and imaginary parts of the quasienergy near EPs of a periodically driven cavity.
We developed an optical detection method referred to as wavefront-angle-resolved band structure spectroscopy to scan the energy band spectrum along the OAM dimension. With the measured complex-energy band structure, the unique spectral features of the EPs, i.e., the existence of Fermi arc, PT symmetry breaking transition, the energy swapping when encircling an EP, and half-integer windings of the paired bands are directly observed. Our results demonstrate the powerful controllability of synthetic OAM dimension and provide a promising platform to investigate singular physics.\\

\begin{figure}[t]
	\centering
	\includegraphics[width=1\columnwidth]{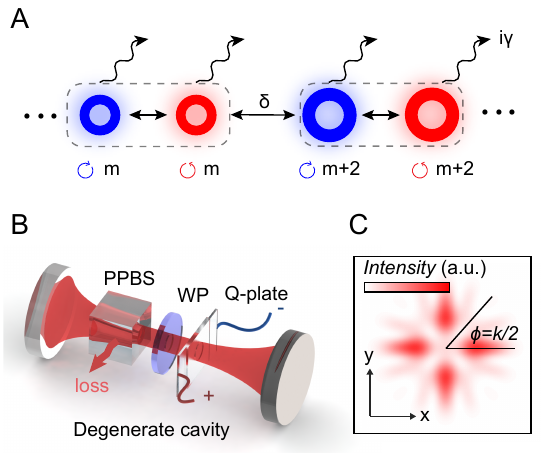}
	\caption{\label{setup}\small{
	\textbf{The synthetic lattice and experimental setup}. \textbf{A}. Schematic of synthetic orbital angular momentum (OAM) lattice. The OAM modes with topological charge $m$ are marked as the arrays of doughnut-shaped rings. The left ($\circlearrowleft$) and right ($\circlearrowright$) circular polarizations are shown in red and blue, respectively. The arrows represent the interaction (straight) and loss (wavy) of optical modes. The coupling strength between different OAM modes controlled by the parameter $\delta$, which serves as another synthetic dimension in momentum space. \textbf{B}. The degenerate cavity accommodates multiple OAM modes with different polarizations. The spin-orbital coupling is achieved by repeatedly passing a wave plate (WP), an anisotropic and inhomogeneous medium (Q-plate) which is controlled by electrical fields. A partially polarized beam splitter (PPBS) is used to introduce the non-Hermitian term.
	\textbf{C}. An example of the average photon transverse distribution from the cavity, which consists of many different polarised OAM modes. The output intensity along quasimomentum $k$ is obtained by scanning the wavefront angle of $\phi=k/2$, which is referred to as the wavefront-angle-resolved band structure spectroscopy.}}
\end{figure}

\noindent{\bf \large Results}\\
In our system, the even-order OAM, labeled by discrete even-numbered topological charges $m$ ($m\in2\mathbf{Z}$), is served as a synthetic lattice. The photons with left-hand ($\circlearrowleft$) and right-hand ($\circlearrowright$) circular polarizations carry corresponding spin angular momenta. The coupling between the OAMs and polarizations is introduced by an anisotropic and inhomogeneous medium named Q-plate~\cite{Qp}, which possesses translational symmetry along OAMs since it plays the same operation on all $m$. The colour doughnut-shaped rings in Fig.~\ref{setup}A represent the corresponding polarized OAM modes. To create a large lattice based on OAM modes, a degenerate optical cavity~\cite{caviye4,caviye5,caviye6} as shown in Fig. \ref{setup}B, resonantly supporting multiple optical modes, is used to trap the polarised photons carrying OAMs.
The operation of Q-plate in terms of spin-orbit coupling (SOC) in ‘momentum’ space can be expressed as,

\begin{equation}
J_{Q}(k,\delta)=\left(
\begin{array}{cc}
\cos(\delta/2)  & i\sin(\delta/2)e^{-ik}  \\
i\sin(\delta/2)e^{ik} & \cos(\delta/2)
\end{array}
\right),
\label{Qk}
\end{equation}
where $k$ represents the quasimomentum of the synthetic OAM lattice, and $\delta$ is the optical retardation controlled by the applied electric field (see methods for more details). $\delta\in[-\pi,\pi]$ determines the coupling strength between optical modes, which can be regarded as an additional pseudomomentum in momentum space. 

\begin{figure}[t]
	\centering
	\includegraphics[width=1\columnwidth]{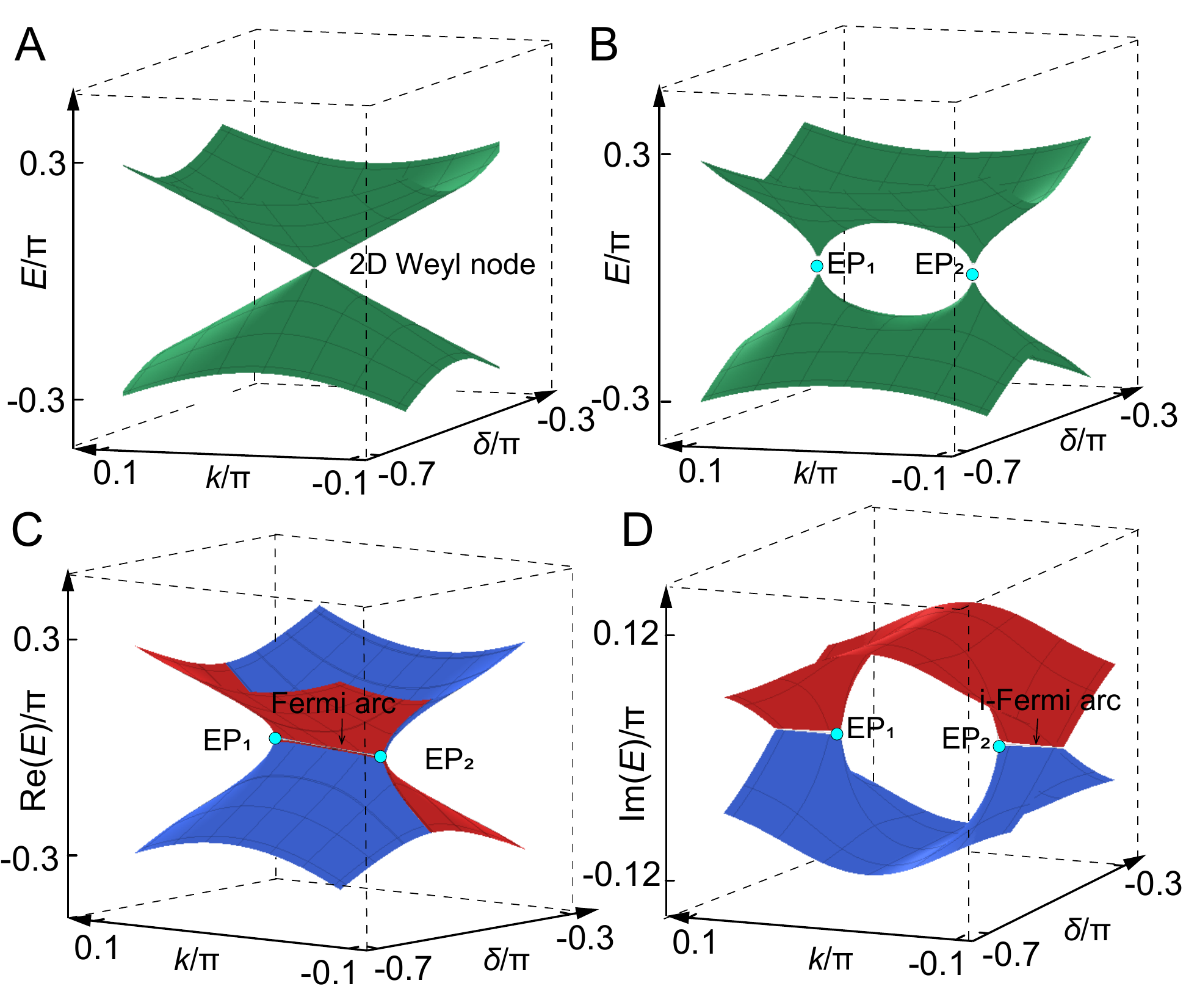}
	\caption{\label{concept}
	\textbf{Theoretical exceptional topological bands by perturbing a Dirac point for $\gamma=0.35$}. \textbf{A}. The band structure near Hermitian Dirac point. \textbf{B}. A pair of exceptional points (EP$_1$ and EP$_{2}$) appear by introducing a non-Hermitian loss. The EPs are labeled in cyan points. The real (\textbf{C}) and imaginary (\textbf{D}) parts of the non-Hermitian bands with the EPs are connected by the Fermi arc and i-Fermi arcs, respectively. The bands of $s=+$ and $s=-$ are labeled in red and blue, respectively.
	}
\end{figure}

\begin{figure*}[t!]
	\begin{center}
		\includegraphics[width=1.8 \columnwidth]{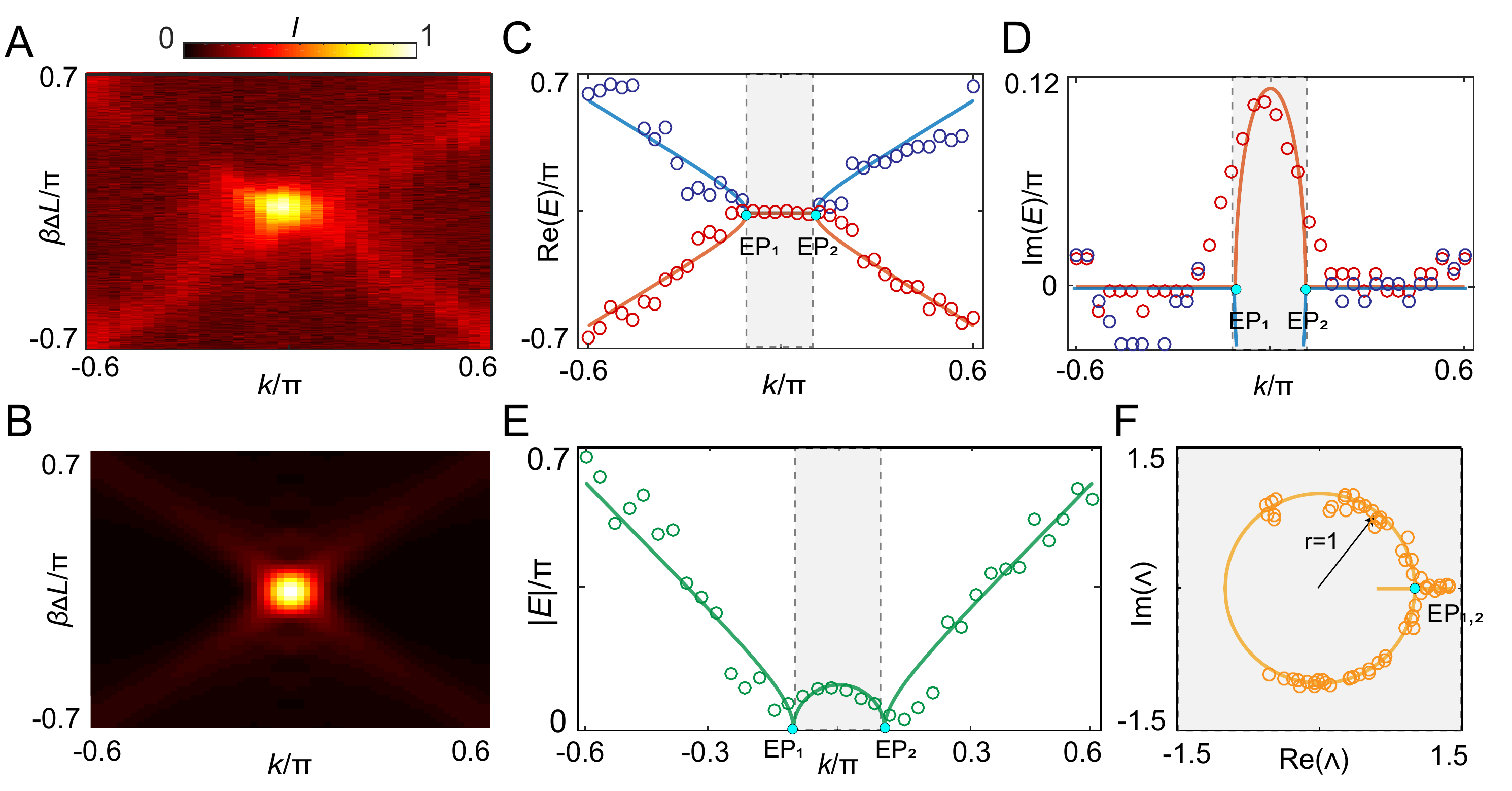}
		\caption{\textbf{Observation of exceptional points and Fermi arcs}. \textbf{A}. The experimentally normalized transmission spectrum $I$ when $\delta=-\pi/2$ and $\gamma=0.35$. \textbf{B}. The theoretical transmission spectrum. \textbf{C}. The real part of complex energy band ${\rm Re}[E_{\pm}(k,-\pi/2)]$ extracted from the cavity detuning $\beta\Delta L$ corresponding to the local maximum transmission. The EPs are labeled in cyan points. The blue and red circles represent the experimental results of ${\rm Re}[E_{+}(k,-\pi/2)]$ and ${\rm Re}[E_{-}(k,-\pi/2)]$ with the solid blue and red lines representing the theoretical predictions, respectively. The bands of $s=+$ and $s=-$ are labeled in red and blue, respectively.
		\textbf{D}. The imaginary part of complex energy band ${\rm Im}[E_{\pm}(k,-\pi/2)]$ extracted from the transmission intensities. The blue and red circles represent the experimental results of ${\rm Im}[E_{+}(k,-\pi/2)]$ and ${\rm Im}[E_{-}(k,-\pi/2)]$ with the solid blue and red lines representing the theoretical predictions, respectively. At the gray region of $k\in [-0.108\pi,0.108\pi]$, only the results of ${\rm Im}[E_{+}(k,-\pi/2)]$ are shown. 
		\textbf{E}. The absolute value of energy $|E_{+}|=\sqrt{{\rm Re}[E_{+}(k,-\pi/2)]^2+{\rm Im}[E_{+}(k,-\pi/2)]^2}$. The green circles and lines represent the experimental and theoretical results, respectively. 
		\textbf{F}. The eigenvalues of the evolution operators $\hat{U}_{\rm NH}$ in the complex plane. The yellow circle and the line crossing the EPs represent the corresponding theoretical predictions, respectively. The yellow circles represent the experimental results. The deviation away from the unit circle indicates the breaking of the PT symmetry.
		}
		\label{EP}
	\end{center}
\end{figure*}

Generally, the EPs can be obtained by an NH type perturbation around a Hermitian Dirac point. As shown in Fig.~\ref{setup}B, a quarter-wave plate (QWP) with the operation $J_{\lambda/4}=e^{i\pi\sigma_{x}/4}$ on polarizations is introduced into the cavity to construct the Dirac point in the Hermitian realm. $\lambda$ represents the wavelength of the input photons. Therefore, the evolution of the photon in one period is $\hat{U}(k,\delta)=J_{Q}(k,\delta)J_{\lambda/4}J_{\lambda/4}J_{Q}(k,\delta)$. Due to the periodicity of the evolution, the effective Hamiltonian of the system can be obtained by $\hat{H}_{\rm eff}(k,\delta)=-i{\rm ln}\hat{U}(k,\delta)$, which can be simplified as a linear function of the Pauli operators $\sigma_{x}$ and $\sigma_{y}$ on the parameter space close to $(k,\delta)=(0,-\pi/2)$. 
The corresponding theoretical band structure is shown in Fig. \ref{concept}A, in which $(0,-\pi/2)$ is the Dirac point.

The non-Hermitian perturbation around the Weyl node is introduced by a non-unitary operator $J_{M}=e^{\gamma\sigma_{x}/2}$, which introduces unbalanced operations on the eigenvectors of $\sigma_{x}$ with eigenvalues of $+1$ and $-1$. $\gamma$ represents the loss control parameter. Experimentally, the operator $J_{M}$ is achieved via a partially polarized beam splitter (PPBS)~\cite{PPBS2} as shown in Fig.~\ref{setup}B. It has a high permeability for horizontally polarized $((\left|\circlearrowright\right\rangle+\left|\circlearrowleft\right\rangle)/\sqrt{2})$ photons. In contrast, only $e^{-\gamma}$ of the photons can be transmitted and the rest will be reflected out of the cavity (lossed) for the vertically polarized $((\left|\circlearrowright\right\rangle-\left|\circlearrowleft\right\rangle)/\sqrt{2})$ photons. The loss effects in the lattice model are denoted as wavy colored in Fig.~\ref{setup}A.

With the non-Hermitian perturbation $J_{M}$, the whole evolution of one round trip in the degenerate cavity is updated to $\hat{U}_{\rm NH}(k,\delta)=J_{Q}(k,\delta)J_{\lambda/4}J_{M}J_{M}J_{\lambda/4}J_{Q}(k,\delta)$. Similarly, we can obtain the effective Hamiltonian $\hat{H}_{\rm NH}(k,\delta)=-i{\rm ln}\hat{U}_{\rm NH}(k,\delta)$ around the Dirac point $(0,-\pi/2)$. The detailed expression of $\hat{H}_{\rm NH}(k,\delta)$ can be found in section I of supplementary material (SM), which has complex eigenvalues. Moreover, a pair of EPs with the coalescence of eigenenergies and eigenvectors appear in the quasimomentum space at $(k,\delta)=[\pm\cos^{-1}(\cosh{\gamma})^{-1},-\pi/2]$. The theoretical simulation result is shown in Fig.~\ref{concept}B. By tuning the parameter $\gamma$, the paired EPs would continuously move in the momentum space and can never vanish except they meet indicating the stability of the non-Hermitian Weyl phase~\cite{NH}. The complex energy bands of $\hat{H}_{\rm NH}(k,\delta)$ are subdivided into the real part (Fig.~\ref{concept}C) and imaginary part (Fig.~\ref{concept}D), respectively, in which the Fermi arc and imaginary Fermi (i-Fermi) arc connecting the EPs are also denoted (see section II of SM for more details).

\begin{figure*}[t]
	\begin{center}
		\includegraphics[width=2 \columnwidth]{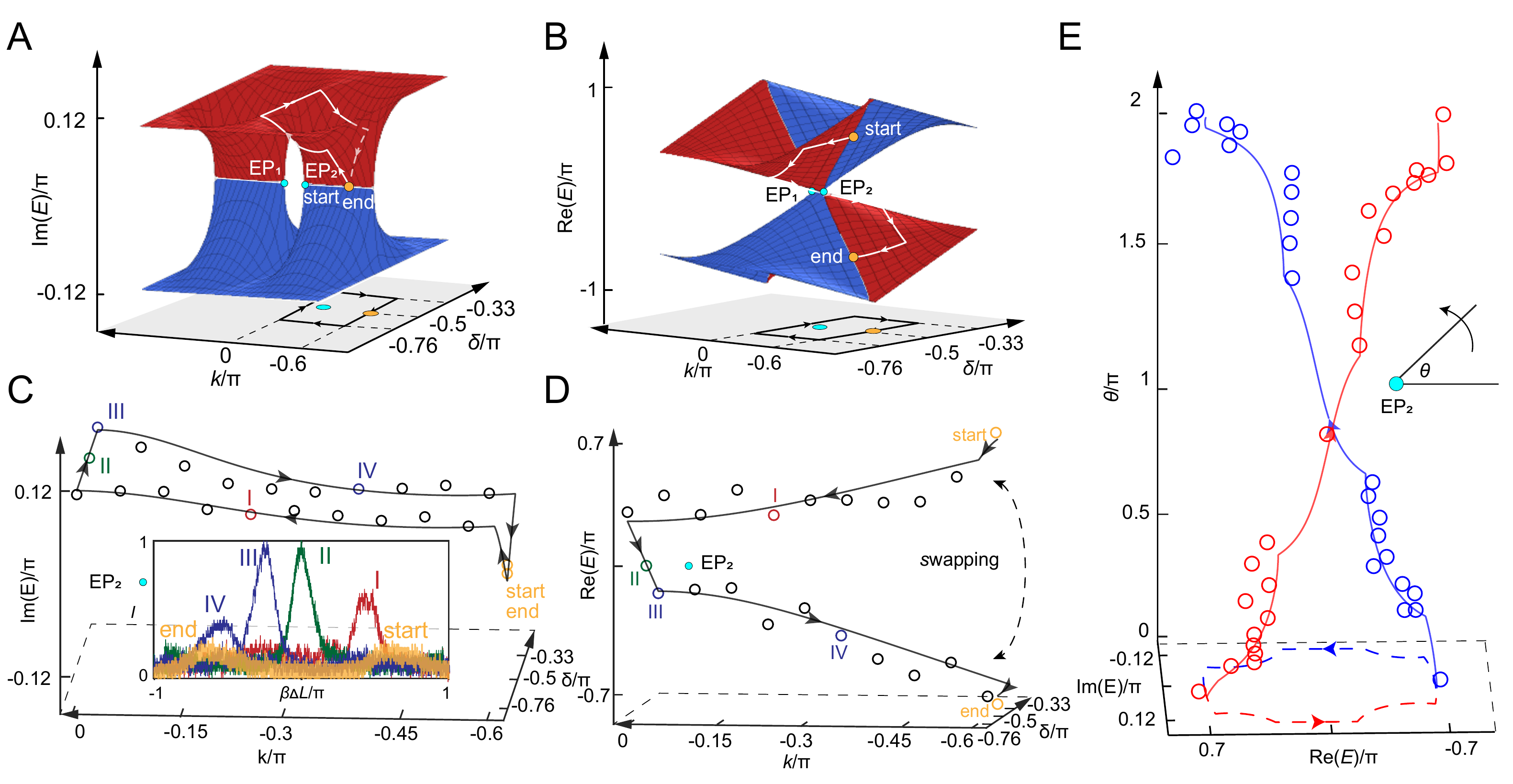}
		\caption{\textbf{The swapping of energy when winding around an EP}.
		\textbf{A}. The winding loop on imaginary energy bands. 
		\textbf{B}. The corresponding winding loop on real energy bands. The winding directions in ($k,\delta$) space around EP$_2$ are: $( -0.6\pi,-0.5\pi)\to$ $(-0.6\pi,-0.76\pi)\to$ $(0,-0.76\pi)\to$ $(0,-0.33\pi)\to$ $( -0.6\pi,-0.33\pi)\to$ $(-0.6\pi,-0.5\pi)$. The bands of $s=+$ and $s=-$ are labeled in red and blue, respectively. The EP$_2$ and start/end points are labeled in cyan and yellow, respectively.
		\textbf{C}. The experimental imaginary part of energy along the loop. Insert: The transmission intensity spectra when ($k,\delta$) are:
		 ($-0.6\pi, -0.5\pi$) (start and end point, yellow),
		\uppercase\expandafter{\romannumeral1} ($-0.25\pi,-0.76\pi$, red), \uppercase\expandafter{\romannumeral2} ($0,-0.5\pi$, green), \uppercase\expandafter{\romannumeral3} ($0,-0.33\pi$, blue), \uppercase\expandafter{\romannumeral4} ($-0.375\pi,-0.33\pi$, purple), which correspond to the color circles in the evolution loops. The observed peaks correspond to $s=+$. The imaginary part of final energy is the same as the start value.
		\textbf{D}. The experimental real part of energy along the loop. 
		The real part of energy swaps from the positive value to the negative value. The circles represent the experimental results and the curves represent the theoretical energy loops.
		\textbf{E}. Windings of the paired bands in the complex-energy plane. The angle $\theta$ around EP$_{2}$ parametrizes the loop and the bands of $s=+$ and $s=-$ are labeled in red and blue, respectively. The circles represent the experimental results, and the arrow solid curves represent the theoretical energy trajectories. The arrow dashed curves are the projection of the energy trajectories.
		}
		\label{wind}
	\end{center}
\end{figure*}

According to the input-output relationship of the cavity, the complex energy band can be directly measured through the transmission intensities of the degenerate cavity, from which the unique features of the EPs would be shown. In the experiment, a continuous-wave laser is used to pump the cavity, and the transmission intensities are recorded by scanning the cavity length $\Delta L$. Since the scanning frequency of cavity length is much smaller than the free spectral range, the system reaches its steady state for each detuning (see section III of SM for more details). As detailedly discussed in sections IV of SM, the normalized transmission intensity $I$ of the system containing all complex energy information is given by
\begin{equation}
\begin{array}{lll}
I=
\dfrac{\Gamma_{0}^{2}}{[\Gamma-{\rm Im}(E_{s}(k,\delta))]^2+[\beta \Delta L-{\rm Re}(E_{s}(k,\delta))]^2}.
\label{T}
\end{array}
\end{equation}
$\Gamma=\Gamma_{0}+\gamma$ and $\Gamma_{0}$ represents the parameter that characterizes the loss of the cavity without the PPBS. $s=\pm1$ corresponds to the two bands of complex energy $E_{s}(k,\delta)$ and we use $s=\pm$ in the rest of the description for simplicity. $\beta=2\pi/\lambda$ represents the wave number.

According to Eq. \ref{T}, when the real part of complex energy (${\rm Re}[E_{\pm}(k,\delta)]$) is equal to the cavity detuning $\beta \Delta L$, the output transmission intensity would reach its local maximum $I_{\rm m}$. At the same time, the imaginary part of complex energy (${\rm Im}[E_{\pm}(k,\delta)]$) can be obtained as $\Gamma-\Gamma_{0}/\sqrt{I_{m}}$. Therefore, the complex band structure can be directly obtained through detecting the transmission intensity spectrum of the cavity.

To obtain the value of $E_{\pm}(k,\delta)$ along $k$, we develop a method referred as the wavefront-angle-resolved band structure spectroscopy. Fig.~\ref{setup}C displays a representative average photon transverse distribution from the cavity. $\phi$ represents the angle parameter of the OAM wavefront in the cylindrical coordinate. The qusimomentum $k$ in the corresponding reciprocal space of the OAM lattice is physically equal to $\phi/2$ and $0<\phi<\pi$ can be regarded as a Brillouin zone. The transmission intensity along the quasimomentum $k$ can be obtained by choosing the photons along angle $2\phi$, from which $E_{\pm}(k,\delta)$ can then be deduced  (see methods and sections V of SM for more details).

The experimental and numerical transmission spectra $I$ with $\delta=-\pi/2$ and $\gamma=0.35$ as a function of $\beta \Delta L$ and $k$ are shown in Fig. \ref{EP}A and Fig. \ref{EP}B, respectively. The experimental results agree with the theoretical predictions.
We find that the maximum transmission intensity appears when $k$ locates around zero ($k\in [-0.108\pi,0.108\pi]$). At the same time, the value of $\Delta L$ is equal to zero, which implies that the real part of the complex energy $E_{\pm}(k,-\pi/2)$ degenerates and becomes zero, denoted as a Fermi arc. We further show the transmission signals summing all $k$ in section VI of SM. The energy band structures can be conveniently engineered by slightly adjusting the cavity (see section VII of SM for more details). 

The experimental result of Re$[E_{\pm}(k,-\pi/2)]$ is shown in Fig. \ref{EP}C. Two local maximal of Re$[E_{\pm}(k,-\pi/2)]$ are found at the region $k\notin[-0.108\pi,0.108\pi]$, which correspond to the two band structures of the spectrum. Blue and red circles represent the experimental results of $E_{+}(k,-\pi/2)$ and $E_{-}(k,-\pi/2)$ with the blue and red solid lines represent the corresponding theoretical predictions, respectively.
Two EPs (EP$_1$ and EP$_2$) denoted as the cyan points appear at $k=\pm0.108\pi$ and are connected by the Fermi arc. 

The unique feature can be seen clearly in the imaginary part of the energy bands (Im[$E_{\pm}(k,\delta)$]). The experimental and theoretical results are shown in Fig.~\ref{EP}D. At the region $k\notin[-0.108\pi,0.108\pi]$, Im[$E_{\pm}(k,-\pi/2)$] degenerates and becomes zero, which means there exists a path connecting two EPs with pure real energy and is denoted as the i-Fermi arc. While for the gray region $k\in [-0.108\pi,0.108\pi]$, because of the overlap of the transmission peaks $I_{\rm m}$ (the degeneracy of the two real bands), we can only get the imaginary energy of the larger peak from Fig. \ref{EP}A, which corresponds to the $s=+$ band. When Re($E_{\pm}(k,-\pi/2)$) is zero, the corresponding imaginary parts are none zero except at EPs, which contribute to the local maximal transmission. The experimental distribution of ${\rm Im}[E_{+}(k,-\pi/2)]$ is a bit wider than theoretical predictions. It is mainly due to the use of multiple optical elements in the cavity, which reduces the fineness. 

With the experimentally measured real and imaginary parts of energy for $s=+$, the results of total absolute value, $|E_{+}|=\sqrt{{\rm Re}[E_{+}(k,-\pi/2)]^2+{\rm Im}[E_{+}(k,-\pi/2)]^2}$ are shown in Fig. \ref{EP}E. The green circles and the solid green lines represent the experimental and theoretical results, respectively. The two EPs are directly determined by $|E_{+}|=0$.

By scanning $k$, the pure real energy (white region) transmits to the pure imaginary energy (gray region), which corresponds to the phase transition from PT symmetry to PT symmetry breaking regime. The PT breaking transition points are EPs. The eigenvalues ($\Lambda=e^{iE_{\pm}(k,-\pi/2)}$) of the
PT-symmetric $\hat{U}_{\rm NH}$ locate on a unit circle. While for the $\hat{U}_{\rm NH}$ with PT symmetry breaking, the eigenvalues will locate inside or outside the unit circle. 
The theoretical prediction of $\Lambda$ is represented by the yellow ring with a line crossing at the EPs.
The experimental results given by the yellow circles in Fig. \ref{EP}F agree well with the theoretical results. 

The complex energy of a two-band NH system lies on a two sheeted Riemann surface. As a result, the corresponding eigenenergy will swap as the tuned momenta encircling an EP. The eigenvector, starting in the upper band, will evolve to the lower band after the winding. 
To demonstrate this feature, we stat/end eigenvector of the system with the energy located on the i-Fermi arc when $(k,\delta)=(-0.6\pi,-0.5\pi)$ (yellow point). The momenta $(k,\delta)$ are adiabatically tuned to encircle the EP$_{2}$ with $(k,\delta)=(-0.108\pi,-0.5\pi)$ (cyan point). The evolution loops are denoted as the white lines in Fig.~\ref{wind}A and  Fig.~\ref{wind}B for Im[$E_{\pm}(k,\delta)$] and Re[$E_{\pm}(k,\delta)$], respectively. The energy bands of $s=+$ and $s=-$ are labled in red and blue, respectively. We can find clearly that the projected evolution loops on the $(k,\delta)$ space encircle the EP$_{2}$.

For each transmission spectrum of $(k,\delta)$, there are two peaks corresponding to the two energy bands with $s=\pm$ (see section VIII of SM). They are distinguishable excepted EPs due to the existence of the energy gap. The typical normalized transmission spectra corresponding to the colored points (star/end (yellow), I (red), II (green), III (purple) and IV (blue)) on the evolution loops are shown in the inset of Fig. \ref{wind}C. For clarity, we only continuously trace one of the transmission peaks (corresponding to $s=+$) to determine the variation of the complex energy here.
The experimental imaginary and real energy are denoted as circles, which are shown in Fig. \ref{wind}C and Fig. \ref{wind}D, respectively. The arrow lines represent the theoretical loops. 
The end point separates from the start point with the energy swapping from $|E_{+}(-0.6\pi,-0.5\pi)|$ to -$|E_{+}(-0.6\pi,-0.5\pi)|$ as encircling the EP$_{2}$. 
The observed phenomenon directly demonstrates the unique characteristics of the Riemann surface of the topological energy band. 

Moreover, the singularity of Riemann energy surface permits the half-integer windings of the paired bands in the complex-energy plane. The loops shown in Figs.~\ref{wind}A and B can be parametrized by an angle $\theta\in[0,2\pi]$ around the EP$_{2}$. The corresponding results of $\theta$ in the complex-energy plane are shown in Fig. \ref{wind}E. The circles are experimental results for the paired bands of $s=+$ (red) and $s=-$ (blue). The arrow solid and dashed curves represent the theoretical loop and projection energy trajectories, respectively. It is be seen that the winding number of each band is $1/2$, which defines a topological invariant associated with the NH band structures~\cite{inv}.\\

\noindent{\bf \large Discussion}\\
In summary, we experimentally explore the singular physics of NH systems based on polarized twisted photons in a cavity. The direct observation of the complex energy bands with EPs in the experiment demonstrates the flexibility of the synthetic platform to characterize the singular properties.
Through introducing the synthetic OAM dimension, many more distinguished NH spectra can be investigated, such as the spectra of edge state~\cite{edge,OAM1} and the spectra with gauge fields~\cite{OAM2}. Moreover, as two exceptional points naturally form a point gap exhibiting the non-Hermitian skin effect~\cite{skin}, signatures connected with the accumulation of OAM modes would be interesting to investigate.
On the other hand, the additional parameter extends the abundance of synthetic topology materials. 
Our work connects the topological photonics, singular optics and non-Hermitian physics, opening exciting opportunities to explore the topological properties of NH systems. In addition, the macroscopic cavities with EPs may contribute to newfashioned multi-mode laser cavities~\cite{mullaser1,mullaser2} and to develop highly sensitive laser gyroscopes~\cite{sensing2,sensing4}. In the future, with the introduction of interphoton interactions, we believe synthetic dimensions will have an advantage in simulating numerically difficult large non-Hermitian systems compared with classical computers.\\

\clearpage
\noindent{\bf \large Materials and Methods}\\
\noindent{\bf Operation of Q-plate}\\
In the real space, the operation of Q-plate is described as
\begin{equation}
\begin{aligned}
J_{Q}(\delta)&=&\sum_{m}\cos(\delta/2)(a_{\circlearrowleft,m}^{\dag}a_{\circlearrowleft,m}+a_{\circlearrowright,m}^{\dag}a_{\circlearrowright,m})\\
&&+i\sin(\delta/2)(a_{\circlearrowright,m+2q}^{\dag}a_{\circlearrowleft,m}+\mathrm{h.c.}),
\end{aligned}
\end{equation}
where $\circlearrowleft (\circlearrowright)$ represents the left (right)-circular polarized modes; $m$ is the topological charge number of OAM mode with corresponding twisted wavefront; $a^{\dag}_{\circlearrowleft(\circlearrowright),m}$ ($a_{\circlearrowleft(\circlearrowright),m}$) denotes the corresponding creation (annihilation) operator; $\delta$ is the relative optical retardation between $\circlearrowleft$ and $\circlearrowright$ modes in the Q-plate which can be tuned by the applied electric field. Note that $\delta$ can be regarded as an additional pseudomomentum, which introduces another synthetic dimension in momentum space. 
$q$ is the topological charge number of the Q-plate and $q=1$ in our experiment. As a result, the OAM transition occurs among even order modes. It is worth mentioning that the long-range coupling can be controlled by increasing the parameter $q$.

Since Q-plate has a same operation on different optical modes, the operation $J_{Q}$ possesses translational symmetry operation on $m$. As a result, we can introduce the Bloch mode $\left|k\right\rangle=\sum_{n}e^{-ink}\left|n\right\rangle$ $(n=m/2)$ in momentum space. The operation can be recast in the `quasi-momentum' space as $J_{Q}(\delta)=\int^{\pi}_{-\pi}J_{Q}(k,\delta)dk$, where $J_{Q}(k,\delta)$ has the form of Eq.\ref{Qk}. \\

\noindent{\bf Wavefront-angle-resolved band structure spectroscopy}\\
To obtain the energy in momentum space, we propose a method to isolate the quasimomentum $k$, named wavefront-angle-resolved band structure spectroscopy.
Within the paraxial approximation, the photon wavefunction $|m\rangle$ carrying even-order OAM with the topological charge of $m$ can be approximately expressed as
\begin{equation}
|m\rangle=E_{0}(r,\phi)e^{im\phi},
\end{equation}
in cylindrical coordinate $(r,\phi)$. $E_{0}(r,\phi)$ represents the amplitude. 
For the lattice consisted by OAMs, the reciprocal Bloch state $|k\rangle=\sum_{n=-\infty}^{n=\infty}e^{-ink}|n\rangle$, where $n=m/2$ represents the number of lattice sites. So the wavefunction corresponding to $|k\rangle$ can be written as
\begin{equation}
\begin{aligned}
|k\rangle&=\sum_{n=-\infty}^{n=\infty}e^{-ink}|n\rangle\\
&=E_{0}(r,\phi)\sum_{n=-\infty}^{n=\infty}e^{-in(k-2\phi)}\\
&=E_{0}(r,\phi)\delta(k, 2\phi),
\end{aligned}
\end{equation}
which means we can post-select the output state on the basis $|k\rangle\langle k|$ by detecting output photons at the angle $\phi=k/2$ through a diaphragm. \\

\noindent{\bf \large Acknowledgement}\\
We acknowledge the helpful discussion with Ze-Di Cheng. This work was supported by the Innovation Program for Quantum Science and Technology (Grants No. 2021ZD0301400, 2021ZD0301200), the National Natural Science Foundation of China (Grants No. 11874343, 61725504, 61327901, 61490711, 11774335, 11821404 and U19A2075), Anhui Initiative in Quantum Information Technologies (AHY060300 and AHY020100), the Fundamental Research Funds for the Central Universities (Grant No. WK2030380017 and WK2470000026, WK5290000003).\\

\noindent{\bf \large Author contributions}\\
M. Y. and Y.-W. L. experimented with the assistant of J.-S. X., Z. H. L. and Z. Y;
H.-Q. Z. and Y.-J. H. contributed to the theoretical analysis with the help of Z. W. Z. and X. X. Z;
J.-S. X, Y.-J. H, C. -F. L. and G. -C. G. supervised the project.
All authors read the paper and discussed the results.\\

\noindent{\bf \large Additional Information}\\
The authors declare no competing financial interests.\\

\noindent{\bf \large Data Availability}\\
All of the data supporting the conclusions are available within the article and the Supplementary Information. Additional data are available from the corresponding authors upon reasonable request.

\end{document}


\title{Supplementary material of ``Realization of exceptional points along a synthetic orbital angular momentum dimension"}

\author{Mu Yang\red{\footnote{These authors contribute equally to this work\label{Contribute}}}}
\affiliation{CAS Key Laboratory of Quantum Information, University of Science and Technology of China, Hefei 230026, People's Republic of China}
\affiliation{CAS Center For Excellence in Quantum Information and Quantum Physics, University of Science and Technology of China, Hefei 230026, People's Republic of China}

\author{Hao-Qing Zhang\textsuperscript{\ref {Contribute}}}
\affiliation{CAS Key Laboratory of Quantum Information, University of Science and Technology of China, Hefei 230026, People's Republic of China}
\affiliation{CAS Center For Excellence in Quantum Information and Quantum Physics, University of Science and Technology of China, Hefei 230026, People's Republic of China}

\author{Yu-Wei Liao\textsuperscript{\ref {Contribute}}}
\affiliation{CAS Key Laboratory of Quantum Information, University of Science and Technology of China, Hefei 230026, People's Republic of China}
\affiliation{CAS Center For Excellence in Quantum Information and Quantum Physics, University of Science and Technology of China, Hefei 230026, People's Republic of China}

\author{Zheng-Hao Liu}
\affiliation{CAS Key Laboratory of Quantum Information, University of Science and Technology of China, Hefei 230026, People's Republic of China}
\affiliation{CAS Center For Excellence in Quantum Information and Quantum Physics, University of Science and Technology of China, Hefei 230026, People's Republic of China}

\author{Zheng-Wei Zhou}
\affiliation{CAS Key Laboratory of Quantum Information, University of Science and Technology of China, Hefei 230026, People's Republic of China}
\affiliation{CAS Center For Excellence in Quantum Information and Quantum Physics, University of Science and Technology of China, Hefei 230026, People's Republic of China}

\author{Xing-Xiang Zhou}
\affiliation{CAS Key Laboratory of Quantum Information, University of Science and Technology of China, Hefei 230026, People's Republic of China}
\affiliation{CAS Center For Excellence in Quantum Information and Quantum Physics, University of Science and Technology of China, Hefei 230026, People's Republic of China}

\author{Jin-Shi Xu}\email{jsxu@ustc.edu.cn}
\affiliation{CAS Key Laboratory of Quantum Information, University of Science and Technology of China, Hefei 230026, People's Republic of China}
\affiliation{CAS Center For Excellence in Quantum Information and Quantum Physics, University of Science and Technology of China, Hefei 230026, People's Republic of China}

\author{Yong-Jian Han}\email{smhan@ustc.edu.cn}
\affiliation{CAS Key Laboratory of Quantum Information, University of Science and Technology of China, Hefei 230026, People's Republic of China}
\affiliation{CAS Center For Excellence in Quantum Information and Quantum Physics, University of Science and Technology of China, Hefei 230026, People's Republic of China}
	
\author{Chuan-Feng Li}\email{cfli@ustc.edu.cn}
\affiliation{CAS Key Laboratory of Quantum Information, University of Science and Technology of China, Hefei 230026, People's Republic of China}
\affiliation{CAS Center For Excellence in Quantum Information and Quantum Physics, University of Science and Technology of China, Hefei 230026, People's Republic of China}

\author{Guang-Can Guo}
\affiliation{CAS Key Laboratory of Quantum Information, University of Science and Technology of China, Hefei 230026, People's Republic of China}
\affiliation{CAS Center For Excellence in Quantum Information and Quantum Physics, University of Science and Technology of China, Hefei 230026, People's Republic of China}
\onecolumngrid
\setcounter{equation}{0}
\setcounter{figure}{0}
\renewcommand{\theequation}{S\arabic{equation}}
\renewcommand{\thefigure}{S\arabic{figure}}

\maketitle	

\newpage
\tableofcontents

\section{Dispersion relations of the cavity}
In a closed cavity, considering photons pass a round trip through the Q-plate, ${\eta/\pi}$-wave plate ($\eta$ represents the phase retardation between ordinary and extraordinary photons, e.g. $\eta=\pi/4$ for a quarter-wave plate) and the partially polarized beam splitter (PPBS), the operation on the photons in the cavity is
\begin{equation}
\begin{aligned}
\hat{U}_{\rm NH}=J_{Q}J_{\eta}J_{M}J_{M}J_{\eta}J_{Q},
\label{ab}
\end{aligned}
\end{equation}
where $J_{Q}$, $J_{\eta}$ and $J_{M}$ represent the operations of Q-plate, $\eta$-wave plate and PPBS, respectively.

To investigate the Non-Hermitian Hamilton with the non-unitary Floquet dynamic process, we introduce the effective Hamiltonian of $\hat{H}_{\rm NH}=i\log\hat{U}_{\rm NH}$. As we define the Bloch modes $|k\rangle=\sum_{n}e^{-ink}|n\rangle$ $(n=m/2)$ along OAM lattices in quasi-momentum space, the Hamiltonian can be written as the spin-orbit coupling (SOC) form, denoted as $\hat{H}_{\rm NH}(k,\delta)=E_{s}(k,\delta)\textbf{n}_{s}(k,\delta) \cdot \boldsymbol{\sigma}$, where   $\textbf{n}_{s}(k,\delta)=[n^{x}_{s}(k,\delta),n^{y}_{s}(k,\delta),n^{z}_{s}(k,\delta)]$ is the complex Bloch vector; $E_{s}(k,\delta)$ represents the quasi-energy and  $s=\pm1$ represent two complex energy bands and we use $s=\pm$ in the rest of description for simplicity; $\boldsymbol{\sigma}=[\sigma_{x},\sigma_{y},\sigma_{z}]$ is the Pauli matrix. Thus, the unitary evolution in quasi-momentum space satisfies
\begin{equation}
\hat{U}_{\rm NH}(k,\delta)=e^{-i \hat{H}_{\rm NH}(k,\delta)}=e^{-iE_{s}(k,\delta)\textbf{n}_{s}(k,\delta)\cdot \boldsymbol{\sigma}}=\cos E_{s}(k,\delta) I - i\sin E_{s}(k,\delta) [\textbf{n}_{s}(k,\delta)\cdot\boldsymbol{\sigma}].
\label{UNH}
\end{equation}

In quasi-momentum space, the operations of Q-plate can be written as
\begin{equation}
J_{Q}(k,\delta)=\left[
\begin{array}{cc}
\cos(\delta/2)  & i\sin(\delta/2)e^{-ik}  \\
i\sin(\delta/2)e^{ik} & \cos(\delta/2)
\end{array}
\right]. 
\label{Qk}
\end{equation}
The action of $\frac{\eta}{\pi}$-wave plate (WP) on spins is
\begin{equation}
J_{\frac{\eta}{\pi}\lambda}=\left[
\begin{array}{cc}
\cos(\eta)  & i\sin(\eta)  \\
i\sin(\eta) & \cos(\eta)
\end{array}
\right]
=e^{i\eta\sigma_{x}},
\label{WPk}
\end{equation}
where $\eta$ is determined by the WP's thickness.
The action of PPBS on spins is 
\begin{equation}
J_{M}=\frac{1}{2}
\left[
\begin{array}{cc}
1  & -i  \\
1 & i
\end{array}
\right]
\left[
\begin{array}{cc}
1  &  \\
 & e^{-\gamma}
\end{array}
\right]
\left[
\begin{array}{cc}
1  & 1  \\
i & -i
\end{array}
\right]
=
e^{-\frac{\gamma}{2}}\left[
\begin{array}{cc}
\cosh(\frac{\gamma}{2})  & \sinh(\frac{\gamma}{2})  \\
\sinh(\frac{\gamma}{2}) & \cosh(\frac{\gamma}{2})
\end{array}
\right]
=e^{-\frac{\gamma}{2}}e^{\frac{\gamma}{2}\sigma_{x}},
\label{Mk}
\end{equation}
where $\gamma=-\ln \sqrt{P}$, and $P$ is the ratio between the transmission and reflection of PPBS with the input state $\frac{\left|\circlearrowright\right\rangle-\left|\circlearrowleft\right\rangle}{\sqrt{2}}$. The $e^{-\frac{\gamma}{2}}$ can be regarded as the additional loss of the cavity (see section \ref{IkT} for details). We redefine the $J_{M}$ as $J_{M}/e^{-\frac{\gamma}{2}}$. 
Compared with Eqs. \ref{ab} and \ref{UNH}, the quasi-energy dispersion relationship can be calculated as
\begin{equation}
\begin{aligned}
 E_{s}(k,\delta)&=s\cosh^{-1}[\cos(2\eta-i\gamma)\cos\delta-\sin(2\eta-i\gamma)\sin\delta\cos k]i. \\
 \label{ee}
\end{aligned}
 \end{equation}
Here, we employed the mathematical relations $\cos(a+ib)=\cosh(b-ia)$ and $\sin(a+ib)=i\sinh(b-ia)$. We can then obtain that
\begin{equation}
\begin{aligned}
    n^{x}_{s}(k,\delta)=& [\cos(2\eta-i\gamma)\cos k \sin\delta+\sin(2\eta-i\gamma)(\cos^2 k\cos\delta+\sin^2k)] / \sin E_{s}(k,\delta), \\
    n^{y}_{s}(k,\delta)=& \sin k [\cos(2\eta-i\gamma)\sin\delta-2\sin(2\eta-i\gamma)\cos k \sin^2 \frac{\delta}{2}]/ \sin E_{s}(k,\delta), \\
    n^{z}_{s}(k,\delta)=&0,
\end{aligned}
\end{equation}
which means that the effective Hamiltonian $\hat{H}_{\rm NH}(k,\delta)$ can be only expanded in terms of $\sigma_{x}$ and $\sigma_{y}$. 

\begin{figure*}[t!]
	\begin{center}
		\includegraphics[width=0.85 \columnwidth]{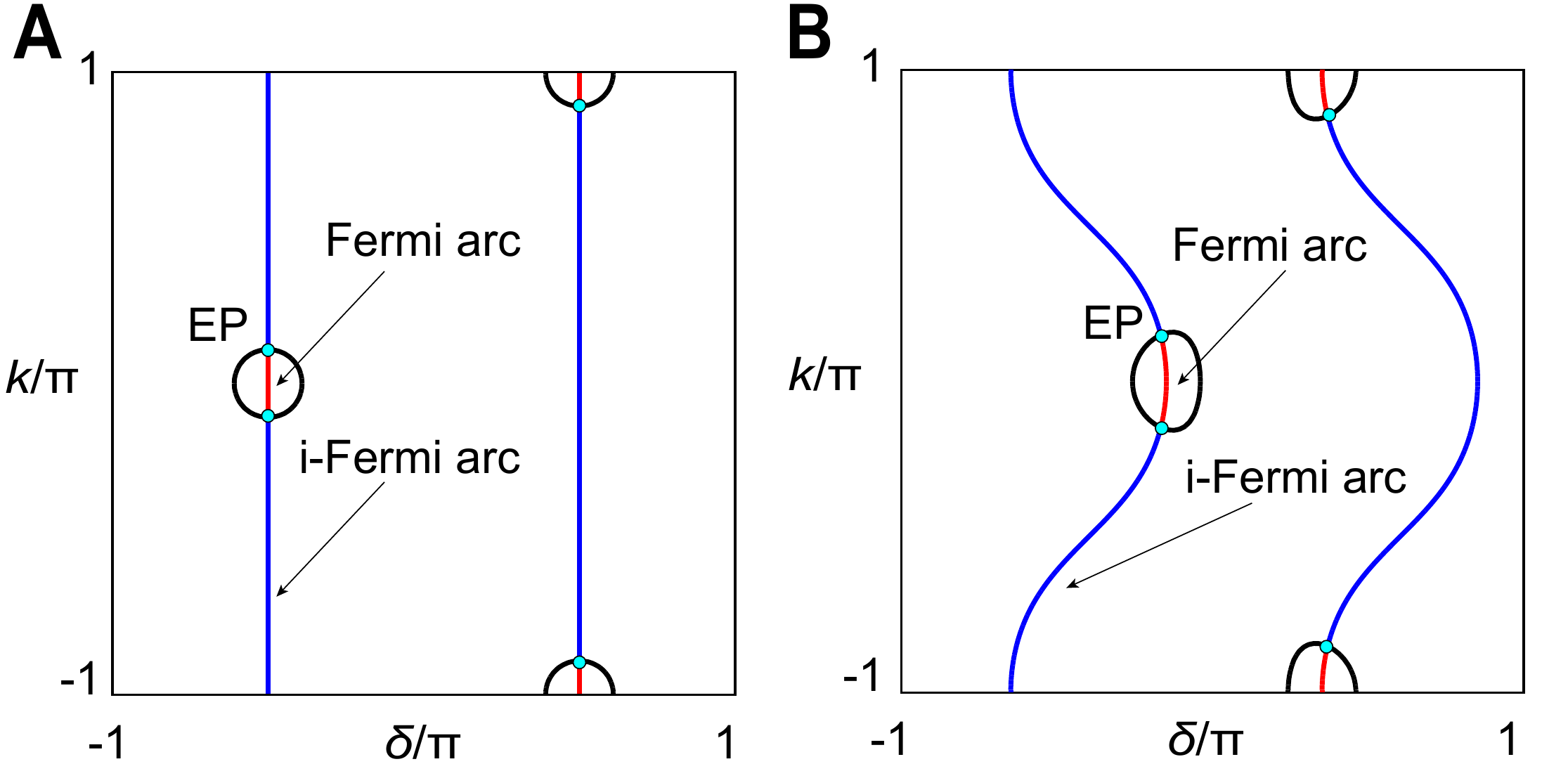}
		\caption{Solutions to $\cosh\gamma(\cos\delta\cos2\eta-\sin2\eta\sin\delta\cos k)=1$ (black lines) and 
    $\sin2\eta\cos\delta+\cos2\eta\sin\delta\cos k=0$ (red and blue lines) when $\gamma=0.35$ form paths in a two dimensional $(k,\delta)$ space as $\eta=\pi/4$ (\textbf{A}) and $\eta=\pi/8$ (\textbf{B}). Exceptional points appear when the two paths intersect, where both equations are satisfied. Exceptional points (cyan points) are connected by Fermi arcs (red lines) and i-Fermi arcs (blue lines).
		}
		\label{arc}
	\end{center}
\end{figure*}

\section{Fermi arcs and exceptional points in the system}
The quasi-energy can form Fermi-arcs in our system, which are unique patterns of a three dimensional Hermition metal. When $\mathrm{Re}[E_{s}(k,\delta)]=0$ is satisfied in the two dimensional momentum space, a line connecting the gap close points appears, denoted as the non-Hermitian Fermi arc. Similarly, for the imaginary part $\mathrm{Im}[E_{s}(k,\delta)]=0$, the imaginary Fermi (i-Fermi) arc also appears. When $\mathrm{Re}[E_{s}(k,\delta)]=0$, the quasimomentum $(k,\delta)$ satisfies:
\begin{equation}
\begin{aligned}
    \cosh\gamma(\cos\delta\cos2\eta-\sin2\eta\sin\delta\cos k)&<1, \\
    \sin2\eta\cos\delta+\cos2\eta\sin\delta\cos k&=0.
    \label{f1}
\end{aligned}
\end{equation}
and $\mathrm{Im}[E_{s}(k,\delta)]=0$ for
\begin{equation}
\begin{aligned}
    \cosh\gamma(\cos\delta\cos2\eta-\sin2\eta\sin\delta\cos k)&>1, \\
    \sin2\eta\cos\delta+\cos2\eta\sin\delta\cos k&=0.
    \label{f2}
\end{aligned}
\end{equation}
It's clearly shown that the expressions for $\mathrm{Re}[E_{s}(k,\delta)]=0$ or $\mathrm{Im}[E_{s}(k,\delta)]=0$ have one free parameter and represents lines in the two dimensional momentum space, as shown in Fig. \ref{arc}. 

When the first inequations in Eq. \ref{f1} and \ref{f2} become equations, the lines of both $\mathrm{Re}[E_{s}(k,\delta)]=0$ and $\mathrm{Im}[E_{s}(k,\delta)]=0$ cross (as the cyan points in Fig. \ref{arc}), which form the exceptional points (EPs) satisfying 
\begin{equation}
\begin{aligned}
    \cosh\gamma(\cos\delta\cos2\eta-\sin2\eta\sin\delta\cos k)&=1, \\
    \sin2\eta\cos\delta+\cos2\eta\sin\delta\cos k&=0.
\end{aligned}
\end{equation}

\begin{figure*}[b!]
	\begin{center}
		\includegraphics[width=0.8 \columnwidth]{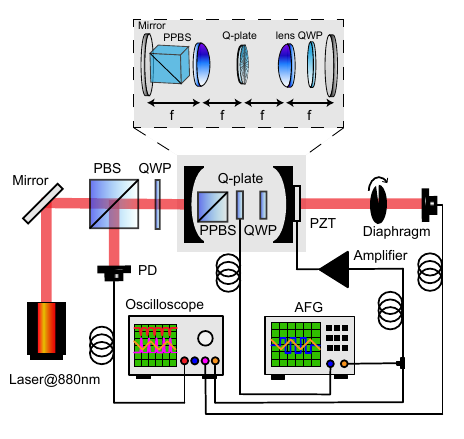}
		\caption{ Schematic diagram of experimental setup. PBS: polarized beam splitter; QWP: quarter-wave plate; PPBS: partially polarized beam splitter; PZT: piezoelectric transducer; PD:  photodetector; AFG: arbitrary function generator.
		}
		\label{setup}
	\end{center}
\end{figure*}

\section{Details of the experimental setup} 
The experimental setup to investigate the band structure of the non-Hermitian system is shown in the Fig. \ref{setup}.  A Gaussian infrared continuous wave (CW) laser with the wavelength at $\lambda=880$ nm is prepared to be left or right circular ($\circlearrowleft$ or $\circlearrowright$) polarized by a polarization beam splitter (PBS) and a quarter-wave plate (QWP) with the optical-axis setting at 45$^\circ$. The laser is used to pump the cavity through a mirror with the ratio between transmission (T) and reflection (R) of T:R=5:95. The uncoupled photons are reflected by the PBS and detected by a photodetector (PD).

The degenerate optical cavity~\cite{cavity1,cavity2} 
consists of two mirrors and two lenses with a focal length of $f=0.1$ m. The two lenses in the cavity form a 4F telescope, allowing any transverse mode that satisfies the self-reproduction condition. The free spectral range (FSR) of the empty cavity is about 375 MHz, and the linewidth is about 13.6 MHz. 

A Q-plate with $q=1$ is placed in the confocal plane of two lenses, and the additional electrostatic field is controlled by an arbitrary function generator (AFG) for adjusting the parameter $\delta$. A $\lambda/4$-wave plate (QWP) behind the Q-plate is used to rotate the polarization. A partially polarized beam splitter (PPBS) in the cavity is used to introduce unbalance loss, where the transmittance is about $50\%$ for the vertically polarized photons ($\frac{\left|\circlearrowright\right\rangle-\left|\circlearrowleft\right\rangle}{\sqrt{2}}$) while the transmittance is $99.9\%$ for the horizontally polarized photons ($\frac{\left|\circlearrowright\right\rangle+\left|\circlearrowleft\right\rangle}{\sqrt{2}}$). A piezoelectric transducer (PZT) is pasted on the output mirror to scan the cavity length $\Delta L$, which is driven by an amplified periodic triangular wave signal generated by the AFG. The frequency of the triangular wave signal is about 40 Hz, which is much less than the FSR of the cavity. As a result, the cavity reaches steady at each length. 

The length of the OAM lattice is limited by the size of the minimal aperture of the system. The transverse radius of the topological charge $m$ of OAM modes is $r_{m}=r_{0}\sqrt{m/2}$~\cite{r}, where $r_{0}$ represents the radius of Gaussian mode ($m=0$). In our experiment, the radius of the Q-plate (minimal aperture) $r_{m}$ is 0.25mm and the radius of pumping Gaussian modes $r_{0}$ is about 80$\mu$m, so the maximal topological charge satisfies $2r_{m}^2/r_{0}^2\approx1.95\times 10^3$.  The length of the synthetic dimension is about $1.95\times 10^3$ and the cavity supporting the even order OAM modes range from $-1.95\times 10^3<m<1.95\times 10^3$.

The transmitted photons out-coupled by the second mirror (T:R=1:99) are spatial filtered by a rotating diaphragm (aperture along $\phi+\Delta\phi$, where $\Delta\phi\approx5^{\circ}$) and then detected directly by a PD. All the signals detected by PDs are displayed at an oscilloscope with a 1 GHz bandwidth to read the energy spectrum column by column along different $\phi$ and $\delta$. 

\section{The input and output of the cavity}\label{IkT}
According to the scattering matrix in Ref. \cite{coupling1,coupling2}, the input and output of the cavity satisfy
\begin{equation}
\left[
\begin{array}{c}
   \phi_{out} \\
   a 
\end{array}
\right]_{\circlearrowleft(\circlearrowright),m}
=
\left[
\begin{array}{cc}
    r & \kappa  \\
    -\kappa^{*} & r^{*}
\end{array}
\right]
\left[
\begin{array}{c}
    \phi_{in} \\ 
    b
\end{array}
\right]_{\circlearrowleft(\circlearrowright),m},
\label{cd}
\end{equation}
where $\kappa=i|\kappa|$ and $r=|r|$ represent the coupling coefficients and $|\kappa|^2+|r|^2=1$. $\phi_{in}$ ($\phi_{out}$) represents the input (output) photonic state. $|a\rangle$ represents the state that has just been coupled into the cavity while $|b\rangle$ is the state that has gone through a round trip.  It is worth mentioning that the there is an phase item $e^{-i\beta L}$ before $|a\rangle$ for the delay in a round trip, where $\beta=2\pi/\lambda$ and $\lambda$ represents the wave length of photons in the cavity. Tush, the input and output relations of the cavity become
\begin{equation}
|b\rangle =\frac{1}{r^{*}}(e^{-i\beta L}|a\rangle+\kappa^{*}|\phi_{in}\rangle),
\label{Sb1}
\end{equation}
and 
\begin{equation}
|\phi_{out}\rangle =\frac{1}{r^{*}} (\kappa e^{-i\beta L}|a\rangle+|\phi_{in}\rangle).
\label{Ob1}
\end{equation}

In a stable cavity, $|b\rangle=e^{-\gamma}\hat{U}_{\rm NH}|a\rangle=e^{-\gamma}e^{-i\hat{H}_{\rm NH}}|a\rangle$, Eq. \ref{Sb1} can be rewritten as
\begin{equation}
\frac{1}{r^{*}}(e^{-i\beta L}|a\rangle+\kappa^{*}|\phi_{in}\rangle)=e^{-\gamma}\hat{U}|a\rangle=e^{-\gamma}e^{-i\hat{H}_{\rm NH}}|a\rangle,
\label{sa1}
\end{equation}
Combining with the Eqs. \ref{Ob1} and \ref{sa1}, we can get the output state as
\begin{equation}
|\phi_{out}\rangle=\frac{1}{r^{*}}|\phi_{in}\rangle-\frac{|\kappa|^{2}/r^{*}}{1-r^{*}e^{-\gamma}e^{i\beta L}e^{-i\hat{H}_{\rm NH}}}|\phi_{in}\rangle.
\label{out1}
\end{equation}
The first term on the right-hand side represents the direct reflection of $|\phi_{in}\rangle$. The second term represents the transmission of the field, and we redefine the $|\phi_{out}\rangle$ as
\begin{equation}
|\phi_{out}\rangle=-\cfrac{|\kappa|^{2}/r^{*}}{1-r^{*}e^{-\gamma}e^{i\beta L}e^{-i\hat{H}_{\rm NH}}}|\phi_{in}\rangle.
\end{equation}
The eigenstates $|\phi_{k,\delta}^{s}\rangle=|\psi_{k,\delta}^{s}\rangle|k\rangle$ of the Hamiltonian $\hat{H}_{\rm NH}$ form a complete basis for expanding $|\phi_{out}\rangle$. Setting the input field as $|\phi_{in}\rangle=|\phi_{in}^{s}\rangle|m_{0}\rangle$, where $|\phi_{in}^{s}\rangle$ and $|m_{0}\rangle$ represent the polarization state and OAM state of input photon, respectively.
The transmission field $|\phi_{out}\rangle$ can be written as 
\begin{equation}
|\phi_{out}\rangle=
\sum_{k,s}\cfrac{-|\kappa|^{2}/r^{*}}{1-r^{*}e^{-\gamma}e^{-i(E_{s}(k,\delta)-\beta L)}}\langle k|m_{0}\rangle \langle\psi_{k,\delta}^{s}|\phi_{in}^{s}\rangle|\psi_{k,\delta}^{s}\rangle|k\rangle.
\end{equation}
Taking $\Delta L$= ${\rm mod}(L,\lambda)$, we define the transmission coefficient $T_{s}(k,\delta)$ as
\begin{equation}
\begin{aligned}
T_{s}(k,\delta)&=\cfrac{-|\kappa|^{2}/r}{1-re^{{\rm Im}[E_{s}(k,\delta)]-\gamma}e^{-i\left\{{\rm Re}[E_{s}(k,\delta)]-\beta \Delta L\right\}}}\\
&\approx\cfrac{-|\kappa|^{2}}{1/r-1+\gamma-{\rm Im}[E_{s}(k,\delta)]+i\left\{{\rm Re}[E_{s}(k,\delta)]-\beta \Delta L\right\}}.\\
\end{aligned}
\end{equation}
Taking $\Gamma=\Gamma_{0}+\gamma=1/r-1+\gamma$, the output intensity $I_{out}=|\phi_{out}|^{2}$ of the output field is
\begin{equation}
\begin{aligned}
I_{out}&=\sum_{k,s}|T_{s}(k,\delta)|^{2}|\langle k|m_{0}\rangle\langle\psi_{k,\delta}^{s}|\phi_{in}^{s}\rangle|^2\\
&=\sum_{k,s}\cfrac{|\kappa|^{4}}{|\Gamma-{\rm Im}[E_{s}(k,\delta)]|^2+|{\rm Re}[E_{s}(k,\delta)]-\beta \Delta L|^2}|\langle k|m_{0}\rangle\langle\psi_{k,\delta}^{s}|\phi_{in}^{s}\rangle|^2.
\end{aligned}
\end{equation}
According to Eq. \ref{ee}, the maximum of ${\rm Im}[E(s,k,\delta)]=\gamma$, thus ${\rm Im}[E(s,k,\delta)]$ is always smaller than $\Gamma$. 
To make $\sum_s|\langle k|m_{0}\rangle\langle\psi_{k,\delta}^{s}|\phi_{in}^{s}\rangle|^2$ be independent of $k$, in our experiment, the input state $|\phi_{in}^{s}\rangle |m_{0}\rangle$ is prepared to the maximally mixed polarization state of $(\left|\circlearrowleft\right\rangle\left\langle\circlearrowleft\right|+\left|\circlearrowright\right\rangle\left\langle\circlearrowright\right|)/2$ with the Gaussian mode $|m_{0}=0\rangle$, the total intensity $I_{\rm out}$ becomes
\begin{equation}
\begin{aligned}
I_{\rm out}
&=\sum_{k,s}\cfrac{|\kappa|^{4}}{|\Gamma-{\rm Im}[E_{s}(k,\delta)]|^2+|{\rm Re}[E_{s}(k,\delta)]-\beta \Delta L|^2}|\langle k|0\rangle|^2(|\langle\psi_{k,\delta}^{s}|\circlearrowleft\rangle|^2+|\langle\psi_{k,\delta}^{s}|\circlearrowright\rangle|^2)\\
&=\sum_{k,s}\cfrac{|\kappa|^{4}}{|\Gamma-{\rm Im}[E_{s}(k,\delta)]|^2+|{\rm Re}[E_{s}(k,\delta)]-\beta \Delta L|^2}.
\label{Iout}
\end{aligned}
\end{equation}
For given $s$ and $\delta$, the normalised transmission intensity $I$ after post selection on the basis $|k\rangle\langle k|$ becomes
\begin{equation}
\begin{aligned}
I&=\cfrac{\Gamma_{0}^{2}}{|\kappa|^4}\sum_{k^{'}k^{''}}\langle k^{'}|k\rangle\langle k|\langle \psi_{k^{'},\delta}^{s}|T_{s}(k^{'},\delta)^{*}T_{s}(k^{''},\delta)|\psi_{k^{''},\delta}^{s}\rangle |k^{''}\rangle\\
&=\cfrac{\Gamma_{0}^{2}}{|\kappa|^4}\sum_{k^{'}k^{''}}|T_{s}(k,\delta)|^{2}\langle\psi_{k^{'},\delta}^{s}|\psi_{k^{''},\delta}^{s}\rangle\delta(k,k^{'})\delta(k,k^{''})\\
&=\cfrac{\Gamma_{0}^{2}}{|\kappa|^4}|T_{s}(k,\delta)|^{2}\\
&=\cfrac{\Gamma_{0}^{2}}{|\Gamma-{\rm Im}[E_{s}(k,\delta)]|^2+|{\rm Re}[E_{s}(k,\delta)]-\beta \Delta L|^2},
\label{ik}
\end{aligned}
\end{equation}
which illustrates the distribution of complex energy $E_{s}(k,\delta)$. By scanning $k$ and $\delta$, the energy band can be directly measured. Specifically, the real part of the complex energy can be read out from the position of the local maximal transmission peak corresponding to ${\rm Re}[E_{s}(k,\delta)]=\beta\Delta L$, and the imaginary part of the complex energy can be calculated via the local maximal intensity $I_{m}$, denoted as ${\rm Im}[E_{s}(k,\delta)]=\Gamma-\Gamma_{0}/\sqrt{I_{m}}$. In section \ref{exam}, we give an example. 

\begin{figure}[t]
	\begin{center}
		\includegraphics[width=0.85 \columnwidth]{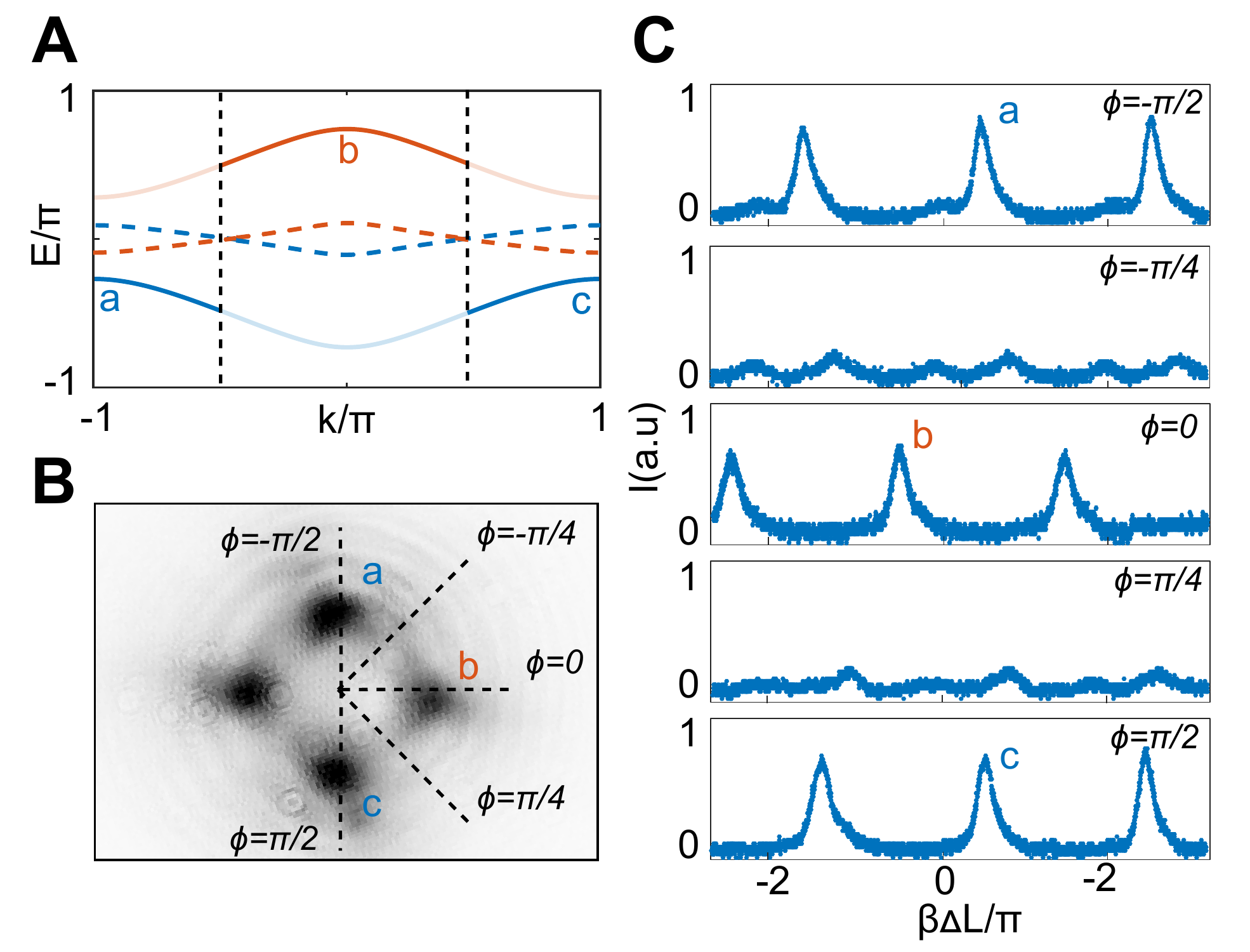}
		\caption{\textbf{The wavefront-angle-resolved band structure spectroscopy.} \textbf{A}. The real (solid lines) and imaginary (dashed lines) energy band $E_{\pm}(k,\pi/2)$ when $\eta=\pi/8$ and $\gamma=0.35$. The imaginary energy bands become zero as $k=\pm\pi/2$ indicated by the black dashed lines. Red and blue lines represent the bands with $s=+$ and $s=-$, respectively. \textbf{B}. The normalized intensity distribution of the output optical beam. \textbf{C}. The transmission intensity spectra along $\beta\Delta L$ of the output optical beam in \textbf{b}. The maximum intensity at the angle of $-\pi/2$, $0$ and $\pi/2$ are labeled as a, b and c, respectively.
		}
		\label{beam}
	\end{center}
\end{figure}

\begin{figure*}[b!]
	\begin{center}
		\includegraphics[width=0.8 \columnwidth]{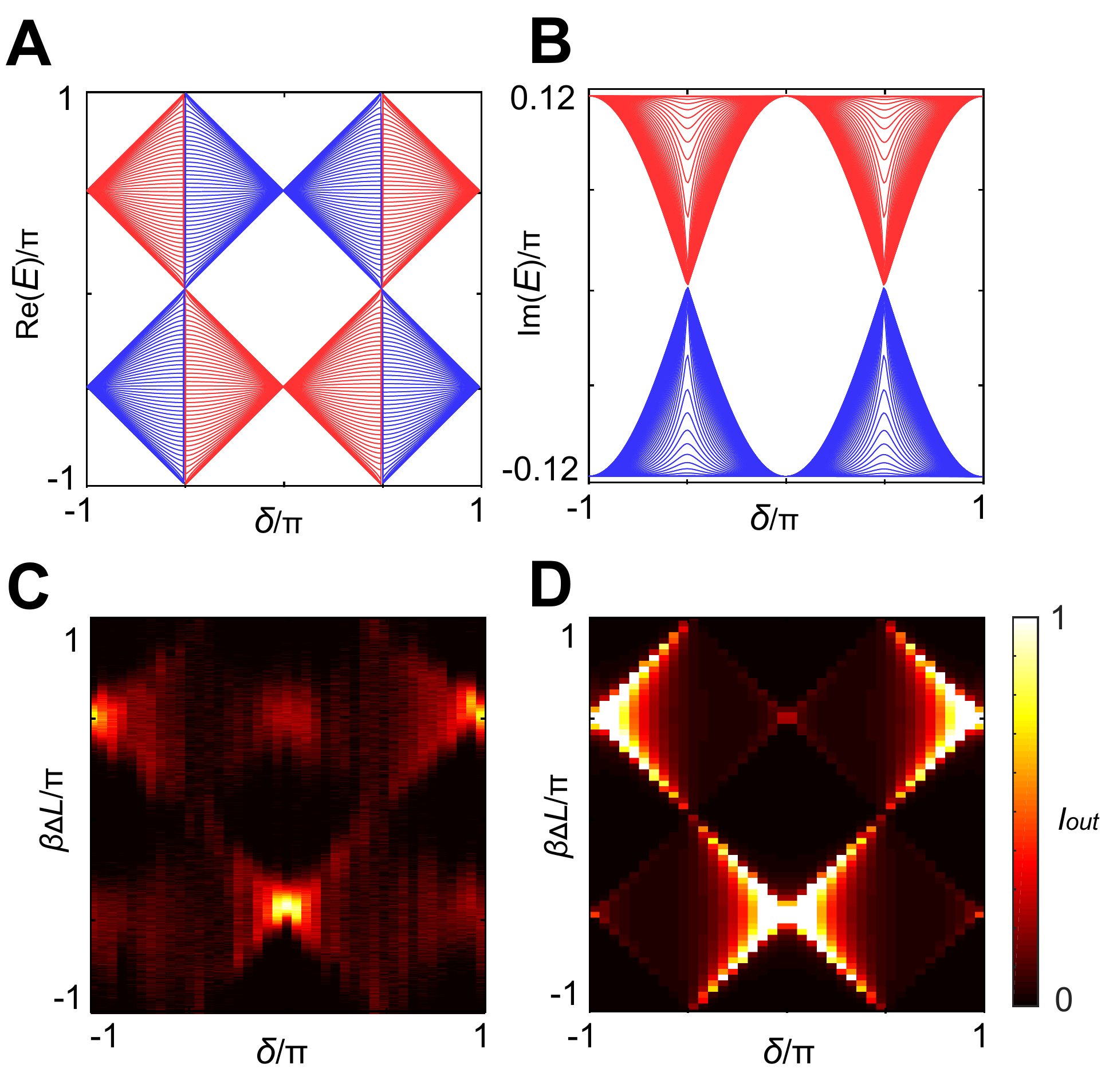}
		\caption{ \textbf{Band structure of all $k,s$ components.} \textbf{A}. The theoretical real band structure vs $\delta$. The bands of $s=+$ and $s=-$ are labeled in red and blue, respectively. \textbf{B}. The theoretical imaginary band structure vs $\delta$.  \textbf{C}. The experimental transmission spectrum $I$. The x and y -axis represent the parameter $\delta$ and the cavity detuning $\beta\Delta L$. \textbf{D}. The numerical transmission spectrum corresponding to \textbf{C}. 
		}
		\label{band}
	\end{center}
\end{figure*}

 \section{An example of wavefront-angle-resolved band structure spectroscopy}\label{exam}
 The method to isolate the quasimomentum $k$ along OAM lattice named wavefront-angle-resolved band structure spectroscopy has been described in section Methods. 
 In this section, we provide an example to demonstrate this method in detail.
 
 The real energy band (solid lines) and imaginary energy band (dashed lines) are shown in the Fig. \ref{beam}A, when $\eta=\pi/8$, $\delta=\pi/2$ and $\gamma=0.35$. Red and blue lines represent the bands with s=+ and s=-, respectively. 
The quasimomentum $k$ is physically equivalent with the angle $2\phi$ of transverse transmission wavefront. Thus, when the real energy $|{\rm Re}[E_{\pm}(k,\pi/2)]|$ reach maximum at $k=0$ $(2\pi)$ and $k=\pm \pi$ (Fig. \ref{beam}A), there are corresponding four highly bright ``petals" in the transverse transmission intensity distribution with the corresponding $\phi=0(\pi)$ and $\phi=\pm \pi/2$, as shown in Fig. \ref{beam}B. 
Besides, the imaginary energy $|{\rm Im}[E_{\pm}(k,\pi/2)]|$ becomes zero as $k=\pm \pi/2$ (Fig. \ref{beam}A). As a result, the output intensity at $\phi=\pm \pi/4$ is close to zero. 

The normalised transmission spectra on the oscilloscope along $\beta\Delta L$ in different $\phi$ are shown in Fig. \ref{beam}C. 
The energy ${\rm Re}[E_{\pm}(k,\pi/2)]$ is physically equivalent to the cavity detuning $\beta\Delta L$ of transmission peaks. According to Eq. \ref{ik}, the output intensity with the positive imaginary energy is higher than that with the negative imaginary energy. 
Thus, as $-\pi/2<\phi<-\pi/4$ and $\pi/4<\phi<\pi/2$, the imaginary energy band with $s=-$ is positive (Fig. \ref{beam}A), the positions of resonant transmission peaks correspond to the real energy band with $s=-$ (Fig. \ref{beam}C). On the other hand, As $-\pi/4<\phi<\pi/4$, the imaginary energy of $s=+$ is positive (Fig. \ref{beam}A), the position of resonant transmission peaks correspond to the real energy band with $s=+$ (Fig. \ref{beam}C). 
 
 See the supplemental video for the detailed progress to detect energy spectrum. 

\section{Identifying Riemann energy surface via the band structure of all $k$ and $s$ components} 
According to Eq. \ref{Iout}, the output intensity $I_{out}$ contains all $k$ and $s$ components. Thus, we can directly obtain the band structure of all $k$ and $s$ components along different $\delta$ to identify the energy bands located on two sheets of Riemann surfaces. 
Here we firstly theoretically investigate the complex band structure that contains all $k$ components along $\delta$, as shown in the Fig. \ref{band}A (real energy band) and Fig. \ref{band}B (imaginary energy band). The real part of energy ${\rm Re}[E_{+}(k,\delta)]>0$ and ${\rm Re}[E_{-}(k,\delta)]<0$ as $-\pi<\delta<-\pi/2$ and $\pi/2<\delta<\pi$, respectively. In contrast, the real part of energy ${\rm Re}[E_{+}(k,\delta)]<0$ and ${\rm Re}[E_{-}(k,\delta)]>0$ as $-\pi/2<\delta<\pi/2$. The energy swapping occurs at the exceptional points (EPs) of $\delta=\pm\pi/2$. This peculiar band structure is due to the energy bands locate on two sheets of Riemann surfaces.

According to Eq. \ref{Iout}, the transmission intensity $I$ as ${\rm Im}[E_{s}(k,\delta)]>0$ would be higher than $I$ as ${\rm Im}[E_{s}(k,\delta)]<0$. In this system, imaginary energy ${\rm Im}[E_{+}(k,\delta)]$ is always positive while ${\rm Im}[E_{-}(k,\delta)]$ is always negative, as shown in Fig. \ref{band}B. So we can judge $s=\pm$ according to the transmission intensity strength $I$. 

The corresponding experimental and numerical transmission spectra along $\delta$ are shown in Fig. \ref{band}C and Fig. \ref{band}D, respectively. The higher output transmission intensity $I_{out}$ corresponding to the band ($s=+$) would swap to the opposite side at $\delta=\pm\pi/2$, which is consistent with the structure of real energy. 

\begin{figure}[t]
	\begin{center}
		\includegraphics[width=0.8 \columnwidth]{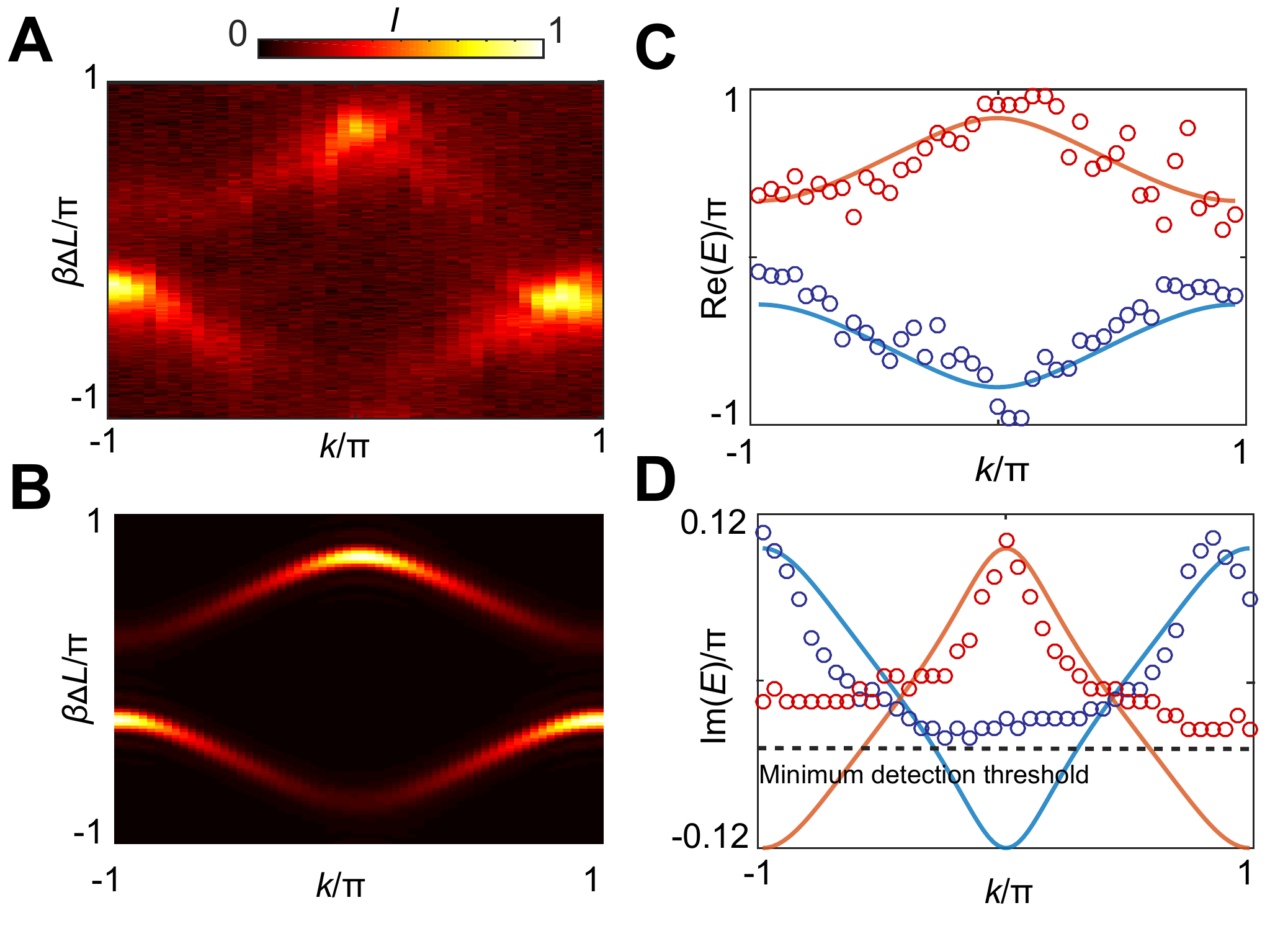}
		\caption{\textbf{The complex energy band when $\delta=\pi/2$, $\eta=\pi/8$ and $\gamma=0.35$}. \textbf{A}. The experimental transmission spectrum $I$. The x and y -axis represent the cavity detuning $\beta\Delta L$ and $k$, respectively. \textbf{B}. The numerical normalised transmission intensity of \textbf{A}. \textbf{C}. The real energy band ${\rm Re}[E_{\pm}(k,\pi/2)]$ extracted from the cavity detuning $\beta\Delta L$ corresponding to local maximum transmission peaks along quasimomentum $k=2\phi$. Red and blue dots are the experimental results of $s=+$ and $s=-$. The solid lines are the theoretical prediction. \textbf{D}. The imaginary energy band ${\rm Im}[E_{\pm}(k,\pi/2)]$ calculated from transmission intensities in \textbf{a}. Red and blue dots are the experimental results of $s=+$ and $s=-$, respectively. The solid lines are the theoretical predictions. Note that ${\rm Im}[E_{\pm}(\pm\pi/2,\pi/2)]=0$. The dashed lines indicate the minimum detection threshold of energy.
		}
		\label{overlap}
	\end{center}
\end{figure}

\section{The complex band structure of $\eta=\pi/8$}\label{soverlap}
Along synthetic dimensions, we can get different band structures by simply changing the parameters of the system. Compared with the photonic crystal system, the synthetic system shows tuning flexibility.
In the main text, the complex bands are introduced with $\eta=\pi/4$. Here we give another example of complex band structure $E_{\pm}(k,\pi/2)$ with $\eta=\pi/8$ without loss of generality. 
Replacing the $1/4$-wave plate in the cavity with $1/8$-wave plate, the imaginary energy bands with $s=+$ and $s=-$ would degenerate at $(k,\delta)=(\pm\pi/2,\pi/2)$, which are the points on the i-Fermi arc. 
The experimental and simulated transmission spectra are shown in Fig. \ref{overlap}A and \ref{overlap}B, respectively. 
Similarly, according to the cavity detuning $\beta\Delta L$ corresponding to local maximum transmission peaks at each $k=2\phi$, we can directly detect the real energy bands ${\rm Re}[E_{\pm}(k,\pi/2)]$, as shown in Fig. \ref{overlap}C. 
According to the intensity of each maximum transmission peak, we can obtain the imaginary energy ${\rm Im}[E_{\pm}(k,\pi/2)]$ as shown in Fig. \ref{overlap}D. 
Note that as the signal to noise ratio of transmitted light intensity is too low, the energy spectrum information of the system cannot be extracted effectively, especially for the negative imaginary energy. Therefore, there is a minimum detection threshold for imaginary energy. This problem may be solved by compensating for the loss of the cavity, such as introducing an optical amplifier into the cavity.

\section{Transmission intensity spectra in momentum space}
\begin{figure}[t]
	\begin{center}
		\includegraphics[width=1\columnwidth]{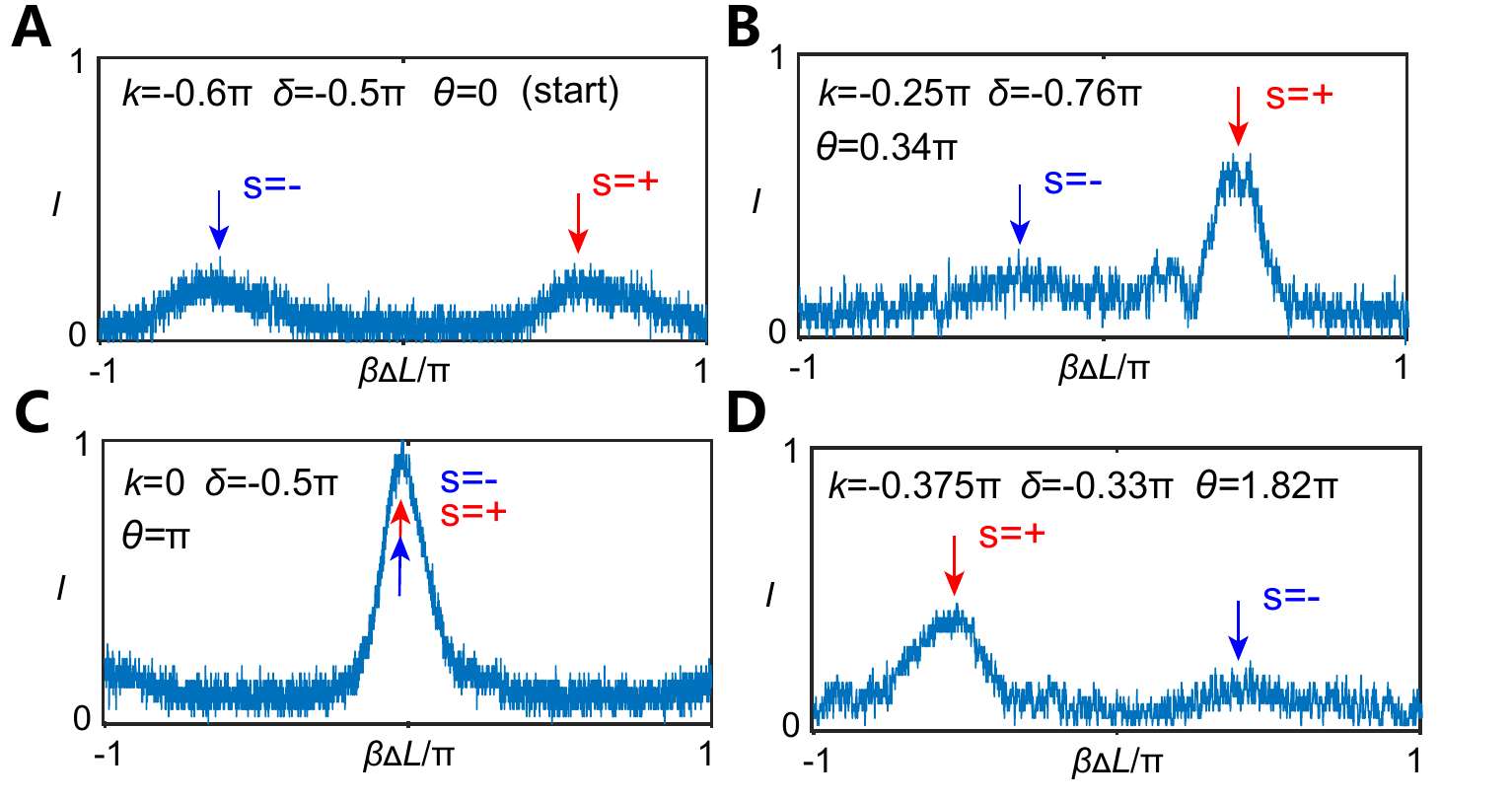}
		\caption{The transmission intensity spectra with different $k$, $\delta$ and $\theta$. The peaks corresponding to the bands of $s=+$ and $s=-$ are marked by red and blue arrows, respectively.}
		\label{spatra}
	\end{center}
\end{figure}

In the study of Riemann energy swapping, we need to trace the peaks of transmission intensity spectra with different parameters. Several typical transmission intensity spectra are shown in Fig. \ref{spatra}. There are two peaks for each transmission intensity spectrum corresponding to the bands of $s=+$ and $s=-$ marked by red and blue arrows. We then obtain the real and imaginary energies according to the cavity detuning $\beta \Delta L$ and the intensities of the peaks, respectively. The two peaks will overlap when the real energies of these two bands are degenerate, as shown in Fig. \ref{spatra}C. We can not read out the imaginary energy in such a situation.
